\documentclass[]{spie}  %>>> use for US letter paper
\usepackage[]{graphicx}
\usepackage{amsmath}
\usepackage{amssymb}

\title{Implementation of kinetics of phase transitions into~hydrocode for simulation of laser ablation} 

\author{Mikhail E. Povarnitsyn, Pavel R. Levashov, and Konstantin V. Khishchenko
\skiplinehalf
Joint Institute for High Temperatures, Russian Academy of Sciences, \\ Izhorskaya~13 bldg~2, Moscow 125412, Russia}

\authorinfo{Further author information: (Send correspondence to M.E.P.)\\
M.E.P.: E-mail: povar@ihed.ras.ru, Telephone: +7 495 484 24 56} 
\begin{document} 
  \maketitle

%%%%%%%%%%%%%%%%%%%%%%%%%%%%%%%%%%%%%%%%%%%%%%%%%%%%%%%%%%%%% 
\begin{abstract}
We model an interaction of femtosecond laser pulses (800~nm, 100~fs, $10^{12}$--$10^{14}$~W/cm$^2$) with metal targets (Al, Au, Cu, and Ni). A detailed analysis of laser-induced phase transitions, melting wave propagation and material decomposition is performed using a thermodynamically complete two-temperature equation of state with stable and metastable phases. Material evaporation from the surface of the target and fast melting wave propagation into the bulk are observed. On rarefaction the liquid phase becomes metastable and its lifetime is estimated using the theory of homogeneous nucleation. Mechanical fragmentation of the target material at high strain rates is also possible as a result of void growth and confluence. In our simulation several ablation mechanisms are observed but the major output of the material is found to originate from the metastable liquid state. It can be decomposed either into a liquid--gas mixture in the vicinity of the critical point, or into droplets at high strain rates and negative pressure. The simulation results correlate with available experiments. 
\end{abstract}

%>>>> Include a list of keywords after the abstract 

\keywords{Laser ablation, nucleation, cavitation, phase transitions}

%%%%%%%%%%%%%%%%%%%%%%%%%%%%%%%%%%%%%%%%%%%%%%%%%%%%%%%%%%%%%
\section{INTRODUCTION}
Laser ablation has become the subject of thorough investigations over the past few decades. Ultrashort laser pulses have already proved their efficiency in machining, microstructuring, synthesis of nanoparticles and medical applications. In metals the conduction band electrons absorb laser energy in a skin layer and then transmit it into the bulk of the target. At the same time the temperature of the lattice rises due to electron--phonon collisions. This two-temperature formulation is used in numerical simulation since the pioneering work of Anisimov \textit{et al.}~\cite{Anisimov}. A lot of theoretical models have been developed since then to describe multistage nonlinear processes of laser--matter interaction. It turned out that simulation depends on many parameters of the model such as frequency of electron-lattice collisions, electron thermal conductivity, optical penetration depth, etc. To achieve good agreement with experiment semi-empirical models with adjustable coefficients are used~\cite{Veysman}.

Simulation of laser-matter interaction is based mainly on continual description~\cite{Colombier, Schafer, Povarnitsyn:PRB:2007} or molecular dynamics~\cite{Ivanov, Perez}. The first approach can deal with real-scale problems but has a difficulty of taking into account the micro-level processes such as void and bubble formation and confluence.  In the second case the dynamics can be studied at the level of individual atoms but the choice and adjustment of an interaction potential requires an extra effort. 

In this study we supplement our hydrodynamic approach with kinetic models of evaporation, nucleation and pressure relaxation to describe main processes of laser ablation of matter. 

%%%%%%%%%%%%%%%%%%%%%%%%%%%%%%%%%%%%%%%%%%%%%%%%%%%%%%%%%%%%%
\section{MODEL}

\subsection{Basic equations}
Our model is based on multi-material Eulerian hydrodynamics with separate descriptions of energy for electron and ion subsystems~\cite{Povarnitsyn:ASS:2007}:
\begin{equation}
\frac{\partial f^{\alpha}}{\partial t}+\nabla\cdot(f^{\alpha} \textbf{u})=\frac{f^{\alpha}\bar{K}_S}{K^{\alpha}_S} \nabla\cdot \textbf{u},
\end{equation}
\begin{equation}
\frac{\partial(f^{\alpha}\rho^{\alpha})}{\partial t}+\nabla\cdot(f^{\alpha} \rho^{\alpha} \textbf{u})=0,
\end{equation}
\begin{equation}
\frac{\partial(\bar{\rho}\textbf{u})}{\partial t}+\nabla\cdot(\bar{\rho}\textbf{u}\otimes \textbf{u}) + \nabla\bar{P}=0,
\end{equation}
\begin{multline}
	\frac{\partial}{\partial t} \left[f^{\alpha}\rho^{\alpha}\left(E^{\alpha}_e+ \frac{|\textbf{u}|^2}{2}\right)\right]+
	\nabla\cdot \biggl[f^{\alpha}\rho^{\alpha}\textbf{u}\Bigl(E^{\alpha}_e + \frac{|\textbf{u}|^2}{2}\Bigr)\biggr]+ \frac{f^{\alpha}\rho^{\alpha}}{\bar{\rho}} \nabla\bar{P}\cdot \textbf{u} = \\ =-\bar{P}_e\frac{f^{\alpha}\bar{K}_S}{K^{\alpha}_S}\nabla\cdot \textbf{u}+Q^{\alpha}_L - 
f^{\alpha}Q^{\alpha}_{ei} +\frac{f^{\alpha}\rho^{\alpha}C^{\alpha}_e}{\bar{\rho}\bar{C}_e} \nabla\cdot\left(\bar{\kappa}_e \nabla\bar{T}_e\right),
\end{multline}
\begin{equation}
\frac{\partial(f^{\alpha}\rho^{\alpha}E^{\alpha}_i)}{\partial t}+\nabla\cdot(f^{\alpha} \rho^{\alpha} E^{\alpha}_i \textbf{u} )=-\bar{P}_i\frac{f^{\alpha}\bar{K}_S}{K^{\alpha}_S} \nabla\cdot \textbf{u}+f^{\alpha}Q^{\alpha}_{ei}.
\end{equation}
Here $f^{\alpha}$, $\rho^{\alpha}$, $K_S^{\alpha}$ and $C_e^{\alpha}$ are the volume fraction, density, isentropic bulk modulus and electron heat capacity of component $\alpha$ respectively; $\textbf{u}$ is the vector of hydrodynamic speed (the same for electrons and ions); $\bar{P}=\bar{P}_e+\bar{P}_i$ is the sum of electron and ion pressures; $\bar{\kappa}_e$ is the effective thermal conductivity; $\bar{T}_e$ is the equilibrium  temperature of electrons; $E^{\alpha}_e$ and $E^{\alpha}_i$ are the specific internal energies of electrons and ions respectively; $Q^{\alpha}_L$  and $Q^{\alpha}_{ei}$ are the source of absorbed laser energy and the electron-ion energy exchange term for  $\alpha$ component respectively. The sum of volume fractions is restricted by the condition $\sum_{\alpha}{f^{\alpha}}=1$, mean density is $\bar{\rho}=\sum_{\alpha}{f^{\alpha}\rho^{\alpha}}$  and effective heat capacity of electrons is $\bar{C}_e=\sum{ \left( f^{\alpha}\rho^{\alpha}C_e^{\alpha}/\bar{\rho}\right)}$. 

For mixed cells we also introduce an effective isentopic bulk modulus and heat conductivity: $1/\bar{K}_S=\sum_{\alpha}{\left(f^{\alpha}/K^{\alpha}_S\right)}$  and $\bar{\rho} \bar{C}_e /\bar{\kappa}_e=\sum{ \left( f^{\alpha}\rho^{\alpha}C_e^{\alpha}/\kappa^{\alpha}_e \right) }$, where $\kappa^{\alpha}_e$ is the electron heat conductivity of  $\alpha$ component.  

Initially, the target occupies the half-space $x \geq 0$. The laser pulse has a Gaussian profile and absorbed energy is estimated 
as $I(x,t)=I_L(1-R)\exp[-\ln(16)(t-t_0)^2/\tau_L{}^2-x/\lambda_{{opt}}]$, 
where $I_L$ is the peak intensity of the laser radiation, $R$ is the reflectivity coefficient,  $\tau_L$ is the full width at half maximum of the pulse and $\lambda_{{opt}}$ is optical penetration depth.  
Absorbed power density is $Q_L=I(x,t)/\lambda_{{opt}}$, and its redistribution between materials in mixed cells is proportional to their mass fractions, $Q^{\alpha}_L = ({f^{\alpha} \rho^{\alpha}}/\bar{\rho}) Q_L$. The energy exchange between electrons and ions, reflectivity and optical penetration depth are calculated as proposed by Eidemann~\textit{el al}.~\cite{Eidemann}.

%The Eulerian formulation has some advantages over a Lagrangian one. For instance, simple spatial splitting may be applied for multi-dimensional simulation. On the other hand, an interface reconstruction algorithm to treat interfaces and free surfaces must be applied~\cite{Pilliod}.

\subsection{Equation of state}
\begin{figure}
   \begin{center}
%\begin{tabular}{c}
   \includegraphics[width=0.8\columnwidth]{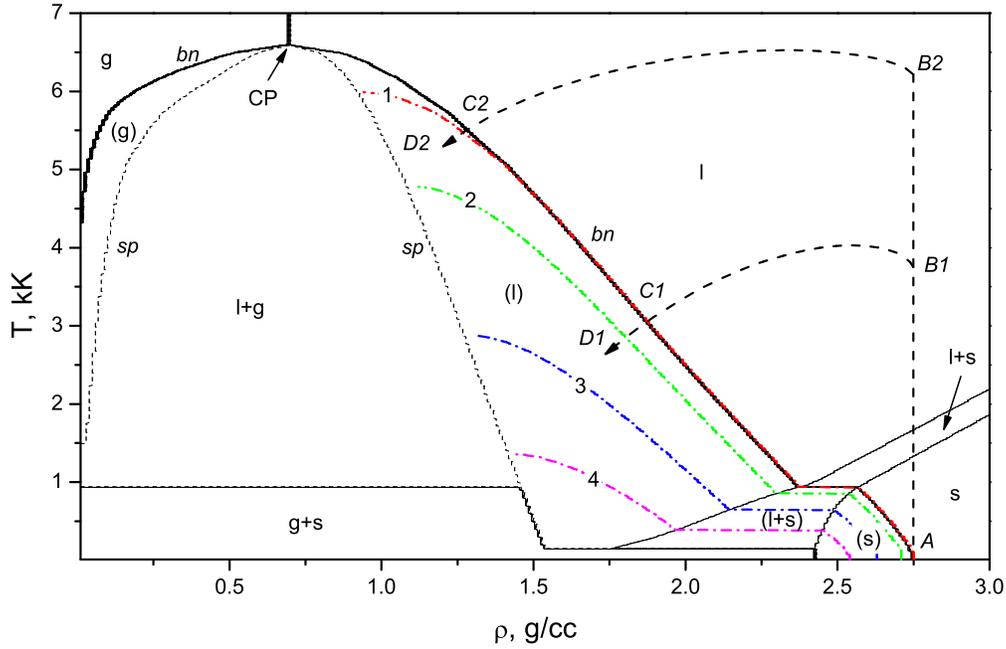}
%\end{tabular}
   \end{center}
   \caption[example] 
%>>>> use \label inside caption to get Fig. number with \ref{}
   { \label{fig:1} 
Phase diagram of aluminum. Here \textit{sp}: spinodal; \textit{bn}: binodal; g: stable gas; l: stable liquid; s: stable solid; l+s: stable melting; l+g: liquid-gas mixture; g+s: sublimation zone; (g): metastable gas; (l): metastable liquid; (l+s): metastable melting; (s): metastable solid; CP: critical point. Dash-dot lines: isobars; 1: 0~GPa; 2: $-1$~GPa; 3: $-3$~GPa; 4: $-5$~GPa.} 
   \end{figure}

We use a semiempirical multi-phase equation of state (EOS) with separate descriptions for subsystems of heavy particles and electrons. The specific Helmholtz free energy has a form $F(\rho, T_i, T_e)=F_i(\rho,T_i)+F_e(\rho, T_e)$, composed of two parts, which describe the contributions of heavy particles and electrons respectively. Here $\rho$ is the material density, while $T_i$ and $T_e$ are temperatures of heavy particles and electrons. The first term $F_i(\rho,T_i) =F_c(\rho) +F_a(\rho, T_i)$, in turn, consists of the electron-ion interaction term $F_c$ (calculated at $T_i=T_e=0$~K) and the contribution of thermal motion of heavy particles $F_a$. The analytical form of $F_i$ has different expressions for the solid $F_i^{s}$, as well as for both liquid and gas phases $F_i^{l}$~\cite{Khishchenko:Elbrus2005:EOS}. Using these thermodynamic functions, the solid, liquid, and gas phases equilibrium boundaries are determined from the equality conditions for the temperature $T_i$, pressure $P_i=\rho^2(\partial F_i/\partial\rho)_{T_i}$, and Gibbs potential $G_i=F_i+P_i/\rho$ of each pair of phases~\cite{LandauLifshits}. The tables of thermodynamic parameters are calculated taking into consideration the information about phase transitions and metastable regions~\cite{Khishchenko1, Oreshkin}. The free energy of electrons in metal $F_e$ has a finite-temperature ideal Fermi-gas form~\cite{LandauLifshits}. Phase diagram of aluminum with stable and metastable states is presented in Fig.~\ref{fig:1}.

\subsection{Treatment of metastable states}
It is known that ultrashort laser pulses result in high-rate isochoric heating (see dashed paths $AB1$ and $AB2$ in Fig.~\ref{fig:1}) followed by the rarefaction of a substance ($B1C1$ and $B2C2$ in Fig.~\ref{fig:1}). On rarefaction the thermodynamic path of substance can cross the liquid branch of the binodal and enter into the metastable region ($C1D1$ and $C2D2$ in Fig.~\ref{fig:1}). The lifetime of this metastable state decreases drastically in the vicinity of the spinodal because of the avalanche-like growth of gas bubbles or voids. The typical space-time scale of this regime is nanometers and picoseconds. This process can be far from equilibrium in terms of hydrodynamics and thus additional kinetic models should be used.

\begin{figure}
   \begin{center}
   \begin{tabular}{c}
   \includegraphics[width=0.8\columnwidth]{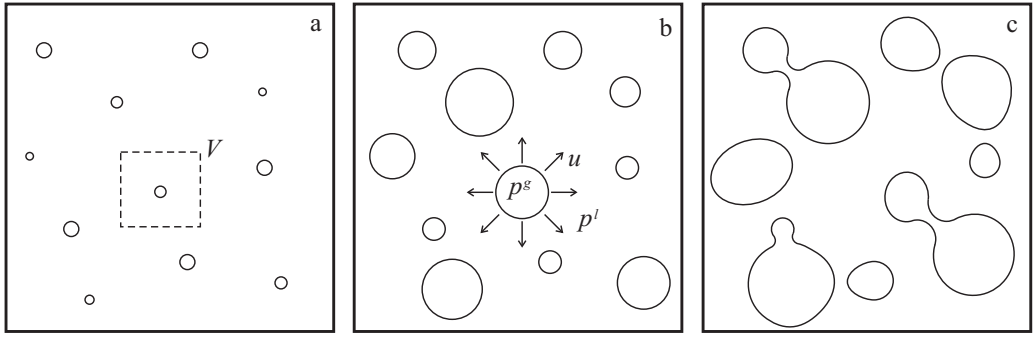}
   \end{tabular}
   \end{center}
   \caption[example] 
%>>>> use \label inside caption to get Fig. number with \ref{}
   { \label{fig:2} 
Three stages of homogeneous nucleation. a: critical size bubble or void formation during the waiting period $\bar{\tau}=(J_1V)^{-1}$; b: void grows with the radial speed $u$; c: void confluence and spallation, typically at $f^{void}\approx 0.15$.} 
   \end{figure}

We consider homogeneous nucleation as the basic mechanism of metastable phase decomposition. It is assumed that nucleation process has three basic stages: (i) appearance of critical size gas bubbles in liquid; (ii) growth of bubbles; (iii) confluence of bubbles and final decomposition, see Fig~\ref{fig:2}. 

The first stage can be described by the critical bubble waiting time $\bar{\tau} = (J_1V)^{-1}$, where $J_1$ is the nucleation rate and $V$ is the volume under consideration~\cite{Skripov}. Stability and growth of a new bubble is governed by the mechanical balance:
\begin{equation}
P^g \geq P^l + 2\sigma/r,
\end{equation}
where $P^g$ and $P^l$ are the pressures in a gas bubble and liquid respectively, $r$ is the radius of the bubble and $\sigma$ is the surface tension. The nucleation rate has a general form~\cite{Skripov}:
\begin{equation}
J_1=Cn\exp(W_c/k_BT),
\end{equation}
where $C\approx10^{10}$~s$^{-1}$ is the kinetic coefficient~\cite{Kagan}, $n$ is the concentration, $W_c$ is the work to create a bubble of a critical radius and $k_B$  is the Boltzmann constant. The work of the critical bubble formation can be derived as~\cite{Skripov}:
\begin{equation}
W_c=\frac{16\pi\sigma^3}{3(P^g-P^l)^2}.
\end{equation}

For bubbles of small radius (nanometers) a contribution of surface curvature into the mechanical balance is important and thus an appropriate wide-range expression for the surface tension should be used. Dependencies known from literature describe surface tension mainly in the vicinity of the critical or triple point, ignoring the large area in a metastable liquid state where pressure is negative, see region below isobar~1 in Fig.~\ref{fig:1}. Keeping in mind the fact that the surface tension must tend to zero on the spinodal and knowing the experimental value at the melting temperature we can extend E\"otv\"os's law for liquid metals into the metastable liquid state:
\begin{equation}
\sigma(T,\rho^l)=\sigma_0\left(\frac{T_c-T}{T_c-T_0}\right)\left(\frac{\rho^l_{bn}(T)-\rho^g_{bn}(T)}{\rho^l_{bn}(T_0)-\rho^g_{bn}(T_0)} \right)^{2/3 }\left(\frac{\rho^l-\rho^l_{sp}(T)}{\rho^l_{bn}(T)-\rho^l_{sp}(T)} \right)^{1/2}.
\end{equation}
Here $T_c$ is the temperature in the critical point, $\sigma_0$ is the surface tension in the triple point at temperature $T_0$, $\rho^l_{bn}(T)$ and $\rho^g_{bn}(T)$ are the densities on the liquid and gas branches of the binodal respectively, and $\rho^l_{sp}(T)$ is the density on the liquid branch of the spinodal.  

On the second stage of the nucleation process the growth of bubbles is governed by the pressure gradient and the speed of the liquid-gas interface can be determined as~\cite{Dergarabedian} 
\begin{equation}
\rho^l r \dot{u} + \frac{3}{2}\rho^l u^2 = P^{g}- P^{l} - \frac{2\sigma}{r}.
\label{drdt}
\end{equation}

On the third stage a confluence of bubbles results in final lost of uniformity by the liquid and formation of stable two-phase liquid-gas state (see Fig.~\ref{fig:2}c). 

\subsection{Model of evaporation}
We use the model of Hertz-Knudsen~\cite{Kelly} to calculate the mass flow of atoms leaving the unit surface during the unit time:
\begin{equation}
J_{evap}=P^g_{bn}\sqrt{\frac{m}{2\pi k_B T}}.
\end{equation}
Here $P^g_{bn}$ is the pressure of the saturated gas (pressure at the binodal) at temperature of condensed phase $T$. Taking into account the back flux in the form 
\begin{equation}
J_{cond}=P^{g}\sqrt{\frac{m}{2\pi k_B T^{g}}}
\end{equation}
we can estimate the total volume fraction of evaporated or condensed substance on the surface $S$ in any volume of interest $V$ during the time $\Delta t$ as $f_{evap}=(J_{evap}-J_{cond}) S \Delta t / (\rho^l V)$. The negative value of $f_{evap}$ indicates that the condensation process predominates.

\subsection{Model of pressure and temperature relaxation}
We adopt the procedures of pressure and temperature relaxation in any multiphase cell of the size $\Delta x$ using the relaxation law $dP/dt = -(P-\bar{P})/\tau_{mech}$, $dT/dt = -(T-\bar{T})/\tau_{therm}$. Here $\tau_{mech} \approx \Delta x/c_s$ and $\tau_{therm} \approx \Delta x^2/(\bar{\kappa}_e/\bar{\rho} \bar{C}_e)$ are the mechanical and thermal relaxation time respectively. This gives us opportunity to describe all stages of nucleation as a balance between mechanical, thermal and chemical processes.
%%%%%%%%%%%%%%%%%%%%%%%%%%%%%%%%%%%%%%%%%%%%%%%%%%%%%%%%%%%%%
\section{RESULTS AND DISCUSSION}
\begin{figure}
   \begin{center}
%\begin{tabular}{c}
    \includegraphics[width=0.8\columnwidth]{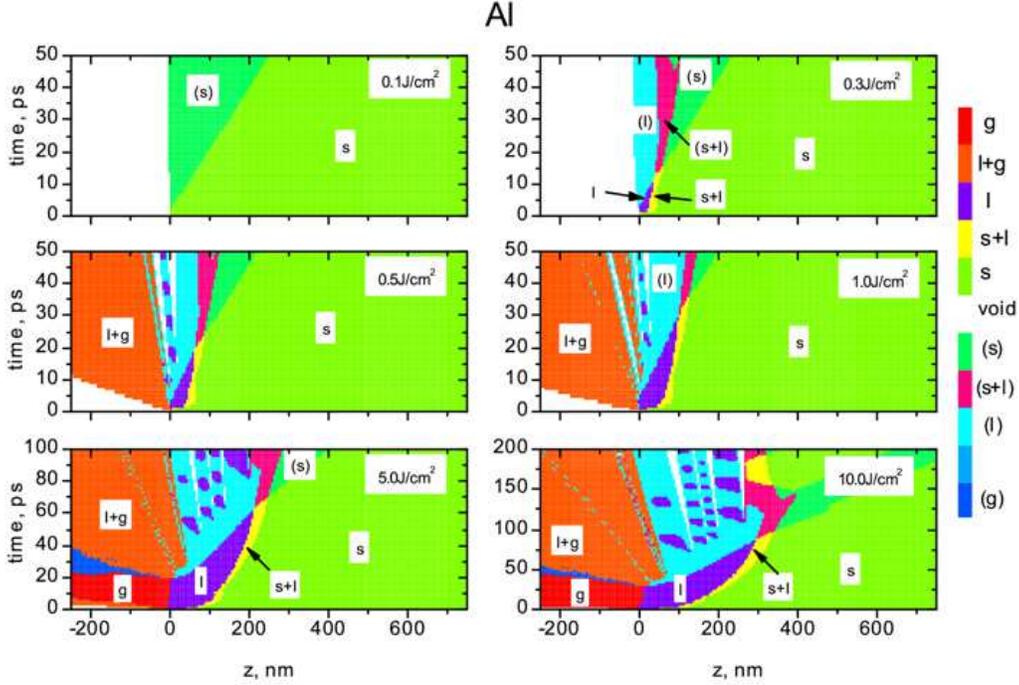}
%\end{tabular}
   \end{center}
   \caption[example] 
%>>>> use \label inside caption to get Fig. number with \ref{}
   { \label{fig:3} Time-space digram of distribution of phase  states for Al.  Here g: stable gas; l: stable liquid; s: stable solid; l+s: stable melting; l+g: liquid-gas mixture; g+s: sublimation zone; (g): metastable gas; (l): metastable liquid; (l+s): metastable melting; (s): metastable solid.} 
   \end{figure}

\begin{figure}
   \begin{center}
%\begin{tabular}{c}
    \includegraphics[width=0.8\columnwidth]{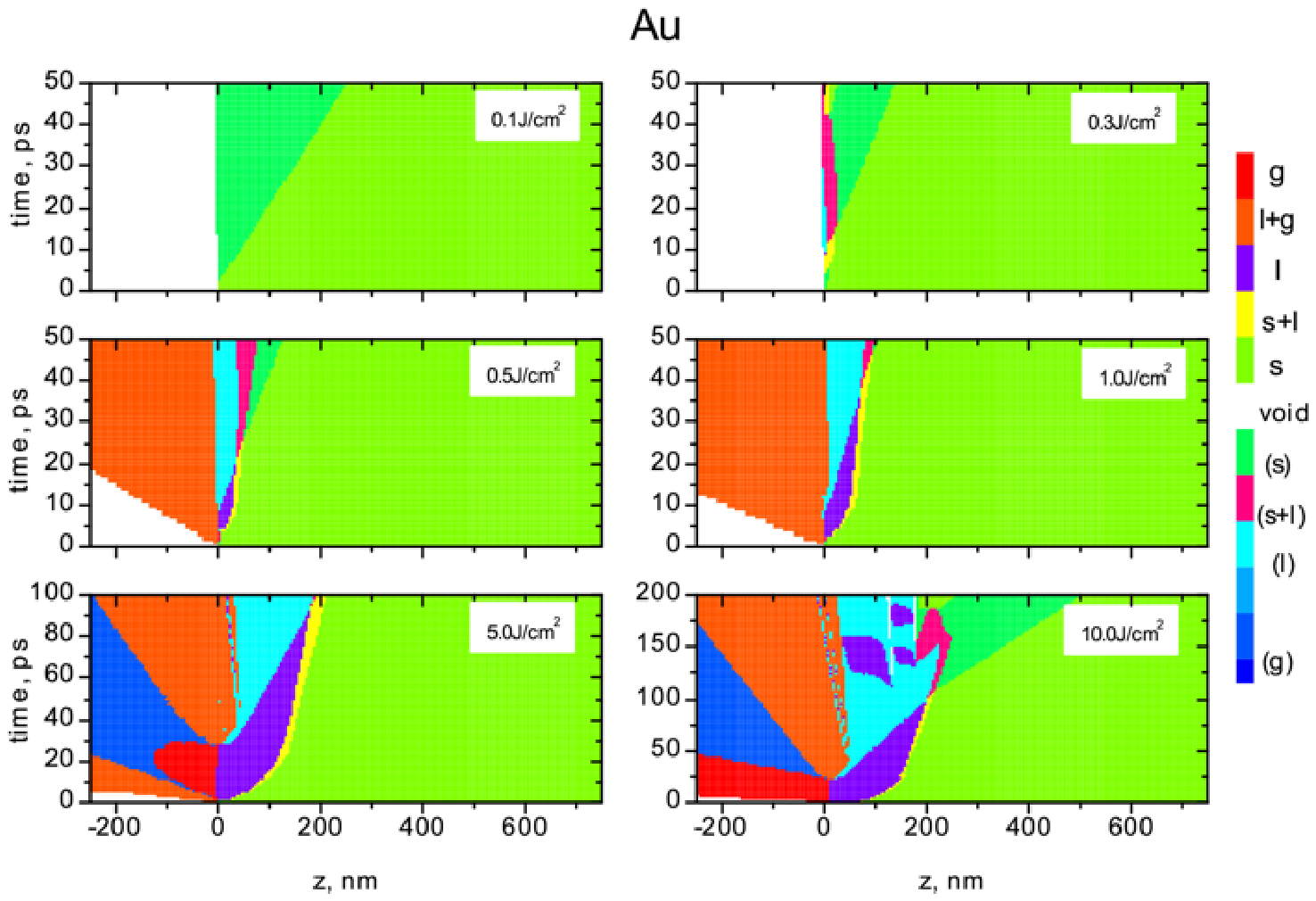}
%\end{tabular}
   \end{center}
   \caption[example] 
%>>>> use \label inside caption to get Fig. number with \ref{}
   { \label{fig:4} The same as in Fig.~\ref{fig:3} but for Au.} 
   \end{figure}

\begin{figure}
   \begin{center}
%\begin{tabular}{c}
   \includegraphics[width=0.8\columnwidth]{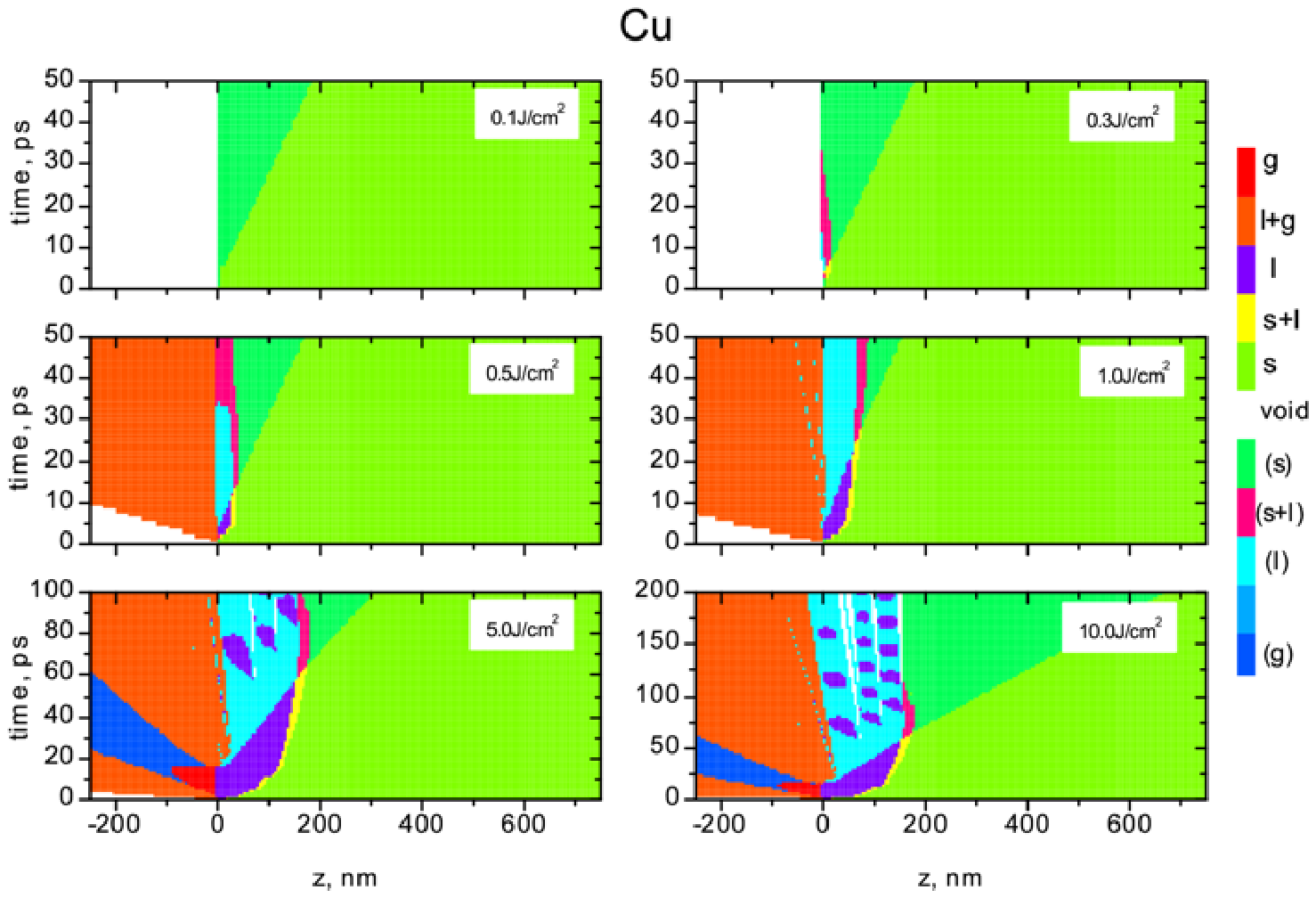}
%\end{tabular}
   \end{center}
   \caption[example] 
%>>>> use \label inside caption to get Fig. number with \ref{}
   { \label{fig:5} The same as in Fig.~\ref{fig:3} but for Cu.} 
   \end{figure}

\begin{figure}
   \begin{center}
%  \begin{tabular}{c}
   \includegraphics[width=0.8\columnwidth]{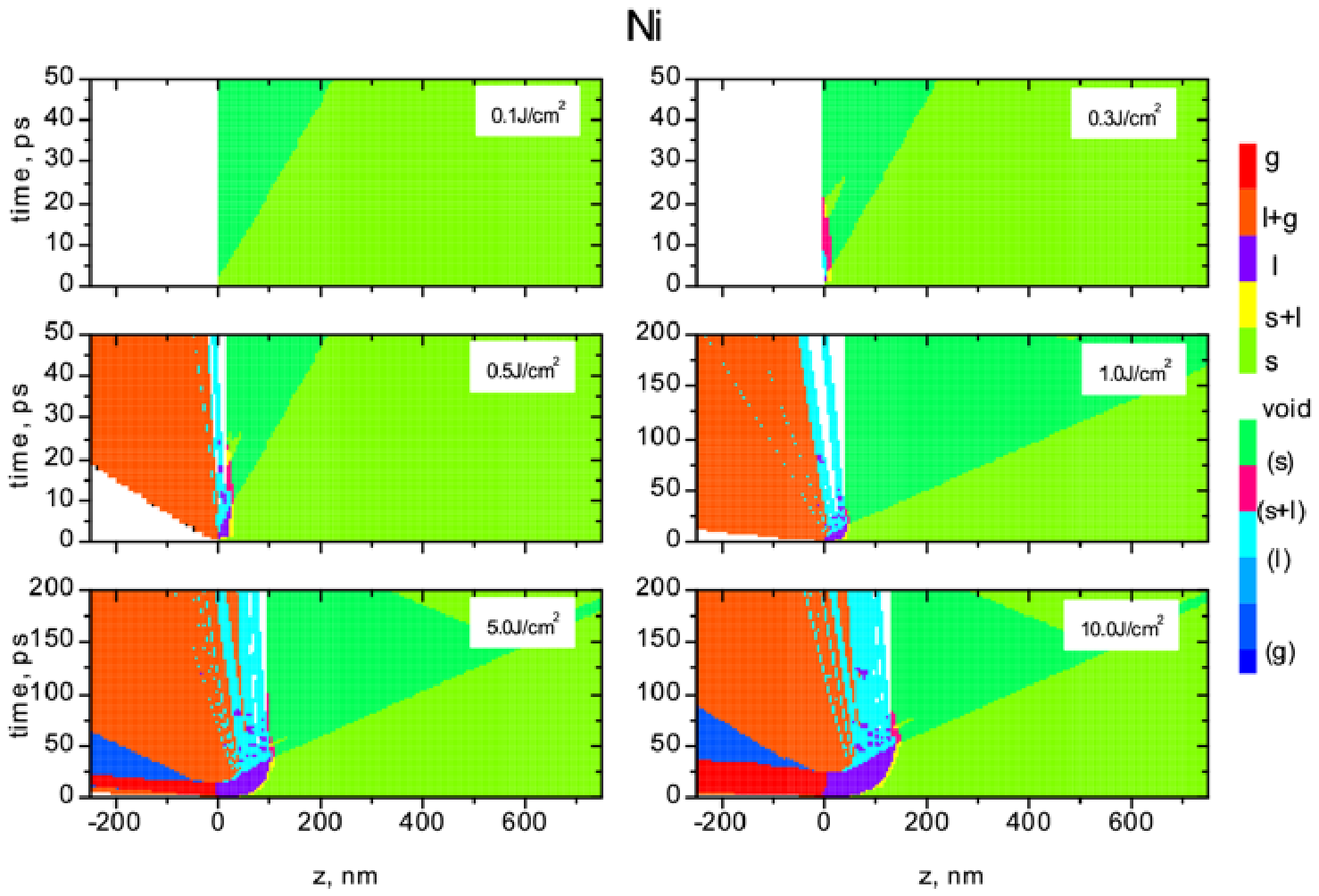} 
%\end{tabular}
   \end{center}
   \caption[example] 
%>>>> use \label inside caption to get Fig. number with \ref{}
   { \label{fig:6} The same as in Fig.~\ref{fig:3} but for Ni.} 
   \end{figure}

We perform simulation of laser-matter interaction for Al, Au, Cu, and Ni and study the dependence of results on the laser fluence. In Figs.~\ref{fig:3}--\ref{fig:6} we present $x$--$t$ diagrams of phase states for these metals after irradiation. At fluences close enough to the ablation threshold ($\sim 0.1$~J/cm$^2$) only mechanical response is observed (formation of a weak shock wave). 

For fluence 0.3~J/cm$^2$, it is seen that melted zone in Al is about 100~nm, whereas for other metals it varies form 20 to 40~nm. The process of melting is reversible as soon as the heat wave moves into the bulk and the temperature on the surface of the target drops. The melted layer becomes thiner at 20~ps for Au, at 10~ps for Cu and Ni, and almost disappears by the moments 50, 30 and 20~ps respectively. For Al the melted layer is thick enough and stops to grow only at about 50~ps. 

For a higher fluence 0.5~J/cm$^2$, the picture of the target response differs a lot. Firstly, an intensive evaporation from the free surface takes place for all metals under consideration and vapor moves away from the target at high speed ($\sim$10~km/s). Then, one can see that the nature of the melted front propagation into the bulk is essentially nonlinear. During the several picoseconds the front goes deep into the bulk overtaking the shock wave. Finally, owing to the intensive tension of the melted layer  on rarefaction a cavitation starts and results in droplet formation (plates in 1D case). Drops have a typical size of tens of nanometers and are ejected at about 1~km/s. 

Further increase of the fluence results in new physical effects. At 5~J/cm$^2$ an intensive evaporation goes in a super-critical regime (above critical point) when transition from the liquid state to the gas one occurs without nucleation. The highest fluence in our simulation is 10~J/cm$^2$ because at bigger laser fluxes ionization effects can not be neglected. 

Simulation shows that ablated material may consist of gas fraction, liquid--gas mixture and liquid droplets. Mass balance between these fractions depends on the laser fluence at a fixed wavelength and pulse duration. Melting and evaporation processes are taken into account based on the enthalpy of melting and evaporation which are determined by the EOS. 

\begin{figure}
  \begin{center}
%\begin{tabular}{c}
      \includegraphics[width=0.8\columnwidth]{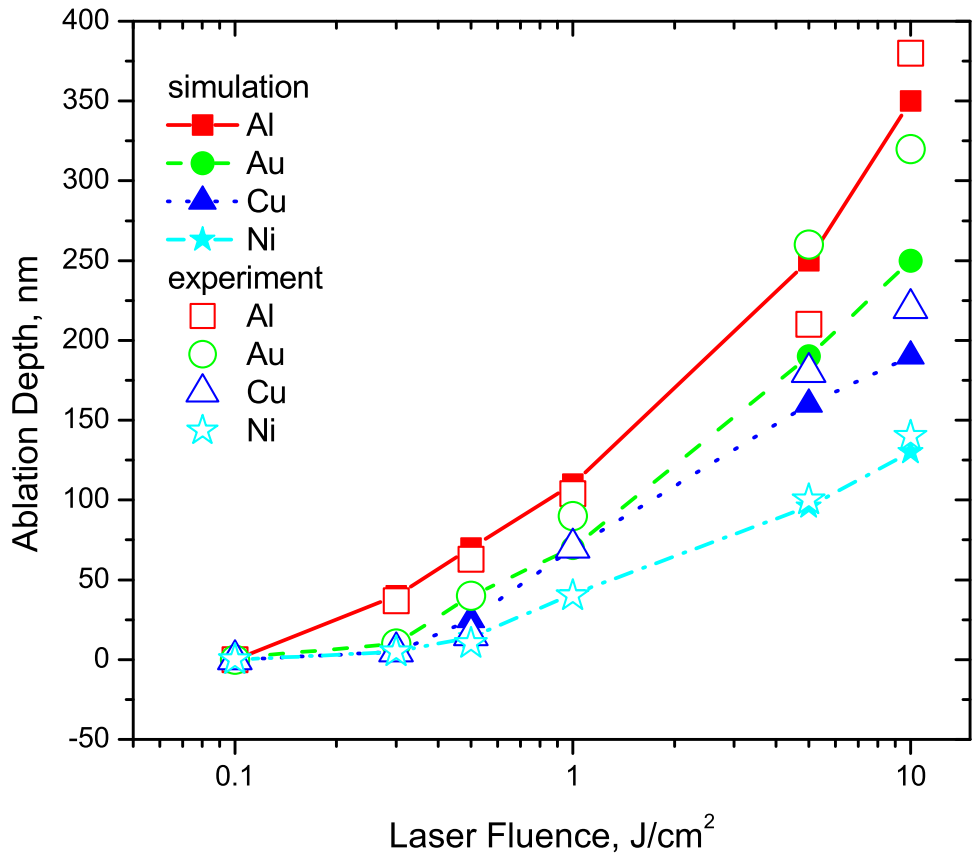}
%\end{tabular}
  \end{center}
  \caption[example] 
  { \label{fig:7} Dependence of ablation depth on laser fluence for Al, Au, Cu, and Ni. Comparison with experiment for Al, Cu, and Ni~\cite{Semerok:2001} and for Au~\cite{Hermann:2008}.}
\end{figure}

We suppose here that the crater formation is governed by the evacuation of evaporated and melted fractions. Under this assumption we can estimate the depth of the crater explicitly as a new interface between the condensed phase and either void or gas. The results of simulation are in a reasonable agreement with experiment. The ablation depth per pulse is presented in Fig.~\ref{fig:7}. Some disagreement is observed for Au and further investigation is required. The possible explanation of underestimation of ablated depth is ignoring the fact that for Au the so called d-electrons can be excited in this regime~\cite{Zhigilei:delectrons:2006}. These electrons can contribute significantly into the heat capacity and electron-phonon coupling and give rise to faster melting and less energy dissipation into the bulk of the target.

\section{CONCLUSION}
The self-consistent model is developed for the description of subpicosecond laser--metal interaction. Phase transitions are taken into account using a multi-phase EOS with separate electron and ion subsystems. Transport properties are modeled using a wide-range model of electron-ion collisions. Cavitation and homogeneous nucleation are the main reasons for material ablation in our model. We observe three mechanisms of laser ablation: (i) direct evaporation from the free surface, (ii) homogeneous nucleation in the neighborhood of the critical point, and (iii) mechanical cavitation in a liquid phase. The mass fraction of ablated material is mainly due to the mechanism (iii) ($\sim 80$\,\%), while effect (ii) produces $\sim10$--15\,\% of the ablated material. Consideration of mechanical effects gives better description of the experimental data.

%%%%%%%%%%%%%%%%%%%%%%%%%%%%%%%%%%%%%%%%%%%%%%%%%%%%%%%%%%%%%
\acknowledgments     %>>>> equivalent to \section*{ACKNOWLEDGMENTS}       
 
This work was supported by the Russian Foundation for Basic Research (projects No.~07-07-00406 and 08-08-01055) and the Council of the President of the Russian Federation for Support of Young Russian Scientists and Leading Scientific Schools (project No.~NSh-6494.2008.2). 
%%%%%%%%%%%%%%%%%%%%%%%%%%%%%%%%%%%%%%%%%%%%%%%%%%%%%%%%%%%%%
%%%%% References %%%%%

%\bibliography{povarefs}   %>>>> bibliography data in report.bib
%\bibliographystyle{spiebib}   %>>>> makes bibtex use spiebib.bst

\hyphenation{Post-Script Sprin-ger}

\end{document}